\documentclass[conference]{IEEEtran}
\IEEEoverridecommandlockouts
\usepackage{cite}
\usepackage{amsmath,amssymb,amsfonts,amsthm}
\usepackage{algorithmic}
\usepackage{graphicx}
\usepackage{textcomp}
\usepackage{xcolor}
\usepackage{subfigure}
\usepackage{breqn}
\usepackage{comment}
\usepackage{bm}
\usepackage[bottom]{footmisc}
\newtheorem{prop}{Proposition}
\newtheorem{cor}{Corollary}
\newtheorem{theorem}{Theorem}
\theoremstyle{plain}

\usepackage[font={small}]{caption}

\usepackage[belowskip=-15pt,aboveskip=0pt]{caption}
\newcommand\numberthis{\addtocounter{equation}{1}\tag{\theequation}}
\def\BibTeX{{\rm B\kern-.05em{\sc i\kern-.025em b}\kern-.08em
    T\kern-.1667em\lower.7ex\hbox{E}\kern-.125emX}}
\begin{document}

\title{The Case for Formation of ISP-Content Providers Consortiums by Nash Bargaining for Internet Content Delivery}
\author{\IEEEauthorblockN{Debasis Mitra}
\IEEEauthorblockA{Columbia University\\
debasismitra@columbia.edu}
\and
\IEEEauthorblockN{Abhinav Sridhar}
\IEEEauthorblockA{Columbia University and Qualcomm\\
as5183@columbia.edu}
\thanks{The authors gratefully acknowledge the benefit of discussions with Qiong Wang (UIUC). 
The work reported in this paper was supported by the National Science Foundation under Grant CMMI-1435378.}}
\maketitle
\setlength{\parindent}{2ex}

\begin{abstract}
The formation of consortiums of a broadband access Internet Service Provider (ISP) and multiple Content Providers (CP) is considered for large-scale content caching. The consortium members share costs from operations and investments in the supporting infrastructure. Correspondingly, the model's cost function includes marginal and fixed costs; the latter has been important in determining industry structure. Also, if Net Neutrality regulations permit, additional network capacity on the ISP's last mile may be contracted by the CPs. The number of subscribers is determined by a combination of users' price elasticity of demand and Quality of Experience. The profit generated by a coalition after pricing and design optimization determines the game's characteristic function. Coalition formation is by a bargaining procedure due to Okada (1996) based on random proposers in a non-cooperative, multi-player game-theoretic framework. A necessary and sufficient condition is obtained for the Grand Coalition to form, which bounds subsidies from large to small contributors. Caching is generally supported even under Net Neutrality regulations. The Grand Coalition's profit matches upper bounds. Numerical results illustrate the analytic results.
\end{abstract}

\section{Introduction}\label{sec1}
The growth of commercial grade Web applications, such as video streaming, has been rapid. Content from providers such as Netflix and YouTube already constitutes a substantial fraction of all Internet traffic, and the trend is strongly upwards. For instance, in the first quarter of 2018, Netflix and YouTube accounted for over 50\% of Internet traffic during peak hours every day, with their respective shares of the downstream traffic being 31.62\% and 18.69\% \cite{b1}. Netflix itself has 125 million subscribers world-wide in 2018 \cite{b2}. This poses well-recognized challenges, and caching (in all its various forms) has been recognized as an important technique to address performance in this environment. See, for instance, T. Leighton's illuminating paper \cite{b3}. Caching to be effective requires cooperation at various levels, ranging from engineering to business, between the players, i.e., Content Providers, the ISP that controls the last mile to the end-user, and the transit ISPs. Lack of such cooperation was plainly visible in 2014 in the national press on the matter of the throttling of Netflix's traffic by Comcast and other ISPs. However, the relationships have turned benign recently \cite{b4}, perhaps in part because of the mutual benefits from the deployment of Netflix's Open Connect caching system \cite{b5}. Besides ISP-CP relationships, other important factors that determine the scale of caching are user behavior and government regulations, notably Net Neutrality. In this paper we take a broad view of the interests and behavior of the afore-mentioned players and factors, with the goal of providing insights into new approaches to very large scale caching for Internet content delivery. \par
It is useful to review the performance bottlenecks that caching addresses. Latency is a major determinant of users' Quality of Experience \cite{b6}. Leighton \cite{b3} argues that the limiting bottleneck, which he calls the "distance bottleneck" and also the "middle mile", is the time spent in traversing the Internet, between the content source and the last mile. It is noted in \cite{b3} that latency and throughput are inversely related. A table in \cite{b3} shows the contrast between the following two connections, (i) local of less than 100 miles, and (ii) cross continental of about 3000 miles. Respective performances are as follows: latency of 1.6 ms vs 48 ms, throughput of 44 Mbps vs 1 Mbps, and download time for 4GB content of 12 min vs 8.2 hrs.\par
A recent paper \cite{b7} models and analyzes related issues based on the Stackelberg leader-follower model of business interaction between ISP and a single CP. The Stackelberg model is well established \cite{b8}. However, we argue that it's use in the context here has several shortcomings: (a) The assumption of the ISP as leader and CP as follower may not be justified currently and, arguably, in the future as well; (b) as a model of business interaction it leaves much to be desired, i.e., comparisons of combined profits of all players in the Stackelberg models to the normative benchmark is highly unfavorable to the former; (c) predictions of effects of Net Neutrality policy are questionable. Let us explain.\par
Susan Crawford \cite{b9} wrote vividly in 2013 about Netflix's fight for survival in the face of the opposition of incumbents, notably Comcast. The picture today is rather different. Netflix's market capitalization is \$ 173 Billion, while Comcast's is \$ 161 Billion. It is positioned today as a leader of the content streaming industry, and indeed all the major streaming video companies are major players with considerable heft. The bargaining power of these players, exemplified by Netflix, is substantial, and certainly in excess of what follower status in the Stackelberg model bestows.\par
The data in \cite{b7} shows that the combined profit of the ISP and CP in the Stackelberg model is substantially less than what may be considered the normative benchmark, i.e., the profit realized in the idealized setting in which the players cooperate in pricing and system design to maximize combined profit. However, no model is proposed for how the normative profit benchmark could be realized in a competitive setting.\par
In the analysis of Net Neutrality, it is proven in \cite{b7} that if regulations prohibit incremental network capacity to the CP's service, then caching collapses. The explanation given is that it is in the ISP's interest to increment the unit price for cache since it increases its profit, even though it is in ever decreasing amounts, and the CP as a follower must accept; however, the impact of such price increases on the CP's profit is increasingly negative, leading it to decrease cache deployment until the null point is reached. Any observer will be justified in thinking that there must be a better way to run a business!\par
There are other important deficiencies in \cite{b7} that are addressed here. The most important of these is the fixed common costs that are incurred by the ISP and CPs. The fixed costs are due to infrastructure, equipment, building, stable and reliable power supply, air-conditioning, repairs, etc. Such costs as distinct from marginal costs that are proportional to the number of subscribers, cache and enhanced network capacity. Fixed common costs are well-known factors in the economics of the communications industry, where they are typically large, and are known to have significant impact on industry structure with respect to barriers to entry, formation of natural monopolies, and regulations; see, for instance, \cite{b14} and \cite{b20}. \par
A related feature of the model here, absent in \cite{b7}, is the presence of multiple, possibly heterogenous CPs. Especially in the presence of large fixed costs, a natural business framework is where these costs can be shared. Indeed, this is key here. \par
We seek an investment and operating framework in which the industry players compete for profit, yet are driven to cooperate out of self-interest, and also where their bargaining powers are not skewed. The framework that we propose is non-cooperative game theory, with Nash bargaining as the basis for coalition formation. More specifically, although the results apply for other protocols, we focus on a protocol with random proposers due to Okada \cite{b18}. The well-known advantage to the first proposer is eliminated and a level bargaining field exists.\par
We give an overview of our results. (i) Our first challenge is to obtain the characteristic function values for any coalition, or subset, of the set of all players, which comprise one ISP and K CPs. These values correspond to the coalition's profit obtained after design optimization. The profit function has the attractive feature of being additive over components corresponding to each CP with the only coupling due to the common fixed cost F. The optimization is with respect to the price (or fee) paid by its subscribers, the cache size and incremental network capacity for each CP. We obtain the optimum price explicitly, and show that the profit function is concave with respect to cache size, which gives a simple condition for its optimum. (ii) For the protocol of random proposers, Okada \cite{b18} obtains a necessary and sufficient condition for the Grand Coalition to form. We obtain a simple equivalent condition for our model. An interpretation of this condition is that subsidies from large profit contributors to small profit contributors may exist in the coalition, but are bounded. Specifically, the gap between the average and the smallest contribution is subject to an upper bound. (iii) The operational implication of this result is that it is necessary to apply a form of Coalition Admission Control for the Grand Coalition to form. (iv) Our analysis and numerical results show that Net Neutrality regulations do not lead to the abandonment of caching. This shows that results to the contrary in \cite{b7} are artifacts of the Stackelberg model. (v) We show that if the Coalition Admission Control condition is satisfied by all the players in the coalition, then it will remain satisfied even if CPs drop out. (vi) We compare the egalitarian payoff to all players, which is implied by the bargaining for the formation of the Grand Coalition, to the Shapley Value payoffs, and find that the non-cooperative foundations yield a rather different cost-sharing formula. (vi) Our numerical results illustrate and corroborate the analysis. An example is given where Net Neutrality regulation has the effect of reducing the number of CPs that may join the coalition. Also, an example is given in which cache and incremental network capacity act as economic substitutes over a wide range of parameter values. (We use the terms "consortiums" and "coalitions" interchangeably, typically in business and analytic contexts, respectively).\par
There is a body of prior work on pricing and economics of Content Delivery Networks (CDNs)[13-15]. CDNs act as intermediaries between ISPs and CPs, and they have had an important role in the Internet. However, in the model considered here this role is taken on by the large CPs. This is best exemplified by Netflix, which earlier had relied on CDNs [16], and now is largely its own CDN. Admittedly it remains to be seen what role CDNs will play in the future. Also, there is considerable prior work on algorithms and deployment of caching [17-18]. Since our focus is elsewhere, we rely on a basic and early model of caching [19].
\section{Models} \label{sec2}
\subsection{Model of System and Subscribers}\label{sec2-1}
Our system model has multiple heterogeneous CPs. Each CP may have its content stored in various geographically dispersed locations, typically in the cloud and at some distance from the ISP's datacenters. Assume that CP k has content of size $\Sigma_{k}(k=1,2,\ldots,K)$.\par
While our interest is in latency (of bits), which is measured in units of time, in the analysis we find it convenient to use the notion of network capacity, which is correlated with throughput and behaves in inverse relationship to latency, and thus has units of bits per unit time. \par
The system that we consider, see Fig.1, has a cache of size $S_k$ dedicated to CP k's content; the cache is located at or close to the ISP's datacenter at the termination point of the last mile. In the event of a cache hit, i.e., the content selection by end-user of CP k happens to be stored in the cache, the total latency experienced by the user is for content delivery over the last mile only, whereas in the event of a cache miss there is additional latency due to traversal of the Internet from content source to the last mile, which is typically large. Let $r_1$ and $r_2$, respectively, correspond to the network capacity in the event of a cache hit and cache miss, so that $r_1 > r_2$. \par
Let $h_k$ denote the cache hit probability. Following Serpanos et. al. \cite{b10}, we assume
\begin{figure}[htp]
    \centering
    \includegraphics[height=3.5cm, width=8.5cm]{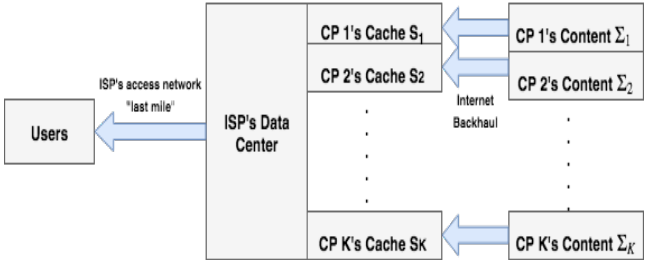}
    \caption{System Model}
    \label{fig:1}
\end{figure}
\begin{equation}
    h_k= \frac{\log{(S_k+1)}}{\log\Sigma_{k}}
\end{equation}
The average network capacity of users of CP k is denoted by $R_k$, where,
\begin{equation}
    R_k=r_1h_k+r_2(1-h_k)
\end{equation}
We consider a second performance-enhancing feature, which is network capacity enhancement of $\beta_{k}$ per user of CP k over the ISP's last mile infrastructure. This enhancement augments $r_1$, the default network capacity for all users. Capacity enhancements need to be negotiated between the CPs and the ISP, and subject to regulations. If the enhancement is enabled then its effect is to modify the average network capacity for users of CP k thus,
\begin{equation} \label{a}
    R_k=(r_1+\beta_k)h_k+r_2(1-h_k)
\end{equation}
\par We adopt the model for user behavior in [7] where the number of subscribers to the service offered by CP k, $n_k$, depends on two major factors, pricing and Quality of Experience (QoE). We let $n_k$ be given in product-form,
\begin{equation}\label{b}
    n_k (p_k, \beta_k, S_k) = D_k (p_k) Q_k (\beta_k, S_k)  \text{ , }      (k=1,2, \ldots, K)             
\end{equation}
Here $D_k (p_k)$ represents the effects of the price for the service, $p_k$, and  $Q_k (\beta_k, S_k)$ represents QoE effects. Specifically, user response to price is based on constant price elasticity of demand [8,20,21]:
\begin{equation} \label{c1}
    D_k=\frac{A_k}{p_k^{\varepsilon_k}}
\end{equation}
where $\varepsilon_k>0$ is the constant elasticity, and $A_k$ is a measure of the potential pool of subscribers. Note that,
\begin{equation}
    \frac{\partial D_k}{\partial p_k}\frac{p_k}{D_k}=-\varepsilon_k,
\end{equation}
i.e., the constant elasticity is the percentage change in demand for an infinitesimal percentage change in the price. We further assume that $\varepsilon_k > 1$, which is consistent with the historical data in \cite{b12} from communications and electric power industries.\par
The function $Q_k$ represents effects of QoE on subscription. Following Reichl at al. \cite{b13}, we let the logarithmic function map QoS to QoE, with QoS given by the average network capacity experienced by users of CP k, $R_k$, i.e.,
\begin{equation}
    Q_k (\beta_k, S_k)  = \log R_k (\beta_k, S_k) 
\end{equation}
Without loss of generality, we choose units such that $r_2$ = 1 in \eqref{a}. Then,
\begin{equation}\label{c}
    R_k (\beta_k, S_k) =  (\delta r + \beta_k) h_k +1  \text{ , }      (k=1,2, \ldots, K)        
\end{equation}                  
where $\delta r = r_1 - r_2 = r_1 - 1 > 0.$
Hence, from \eqref{b} - \eqref{c}, the number of subscribers of CP k,
\begin{equation} \label{d}
    n_k (p_k, \beta_k, S_k) = \frac{A_k}{p_k^{\varepsilon_k}}\log R_k (\beta_k, S_k)
\end{equation}
                              
\subsection{Costs} \label{sec2-2}
We collect here information on costs assumed in the model. Importantly, we have included fixed common costs incurred separately by the ISP and CPs. Let $c$ be the cost incurred by all CPs to support each subscriber, so that the corresponding marginal cost incurred by CP k is $cn_k$. This cost is due to expenses such as billing, authentication, etc. CP k's expense for caching is $c_k^{(S)}$ per unit of cache installed, so that its cache installation cost is $c_k^{(S)}S_k$. It's fixed cost is denoted by $f_k$. \par
The ISP's fixed cost, typically large, is given by $F$. It is tacitly understood that this fixed cost is incurred by the ISP only if an agreement for a coalition to form is reached. We also let $\eta^{(S)}$ and $\eta^{(\beta)}$ respectively denote the unit costs for cache and network capacity incurred by the ISP.
\subsection{Non-Coalitional Business Model} \label{sec2-3}
In this business model, the ISP sells to each CP cache and network capacity at unit prices, and incurs unit costs described above for providing these resources; the unit prices may be specific to the CP. Each CP's revenue comes from fees paid by its subscribers. Let the unit price set by the ISP for the sale of cache and network capacity to CP k be $t_k^{(S)}$ and $t_k^{(\beta)}$, respectively. The profit of the ISP and CP k are given by,
\begin{equation}
    \pi_{ISP}(\bm{p}, \bm{\beta}, \bm{S}) =\sum\limits_{k=1}^{K} \{ (t_k^{(\beta)}-\eta^{(\beta)})\beta_k n_k+ (t_k^{(S)}-\eta^{(S)})S_k \}-F
\end{equation}
\begin{dmath} \label{h}
    \pi_k(p_k,\beta_k,S_k)=(p_k-c-t_k^{(\beta)}\beta_k)n_k-(c_k^{(S))}+t_k^{(S)})S_k-f_k
\end{dmath}
Introducing the expression for $n_k$ in \eqref{d}, we obtain,
\begin{align*} 
    \pi_{ISP}(\bm{p}, \bm{\beta}, \bm{S}) &=\sum\limits_{k=1}^{K} \{ (t_k^{(\beta)}-\eta^{(\beta)})\beta_k \frac{A_k}{p_k^{\varepsilon_k}}\log R_k (\beta_k, S_k) \\
    &+ (t_k^{(S)}-\eta^{(S)})S_k \}-F \numberthis \label{f}
\end{align*}
\begin{dmath} \label{e}
    \pi_k(p_k,\beta_k,S_k)=(p_k-c-t_k^{(\beta)}\beta_k)\frac{A_k}{p_k^{\varepsilon_k}}\log R_k (\beta_k, S_k)-(c_k^{(S))}+t_k^{(S)})S_k \hiderel{-} f_k
\end{dmath}
In the absence of a business agreement between the ISP and the CPs, the cache for CP k, $S_k  = 0$, hence the cache hit probability, $h_k = 0$, and $R_k (\beta_k, 0) = 1, Q_k (\beta_k, 0) = 0$, and consequently  $\pi_{ISP} (\bm{p}, \bm{\beta}, \bm{0}) = \pi_k (p_k, \beta_k, 0) = 0  \text{ , } (k=1,2\ldots,K)$.
\subsection{Coalitional Business Model}
We obtain expressions for the profit functions in the idealized setting of "Integrated Operations" (IO) in which the ISP and CPs jointly make business decisions to maximize their combined profit. Thus in this setting there is no transfer of funds from the CPs to the ISP for cache and network capacity.  We note that the full relevance of the model presented here will become apparent in Secs. \ref{sec4} and \ref{sec5}. \par
The notation that we follow is as follows: the players are indexed by k, the CPs' indices range from 1 to K, and the ISP is indexed K+1. We let the set $\mathbb{K}=\{1,2,\ldots,K\}$ and $\mathbb{K}^{+}=\mathbb{K} \cup \{K+1\}$. We denote any coalition of CPs by $\Theta \subseteq \mathbb{K}$ and the coalition inclusive of the ISP by $\Theta^{+}=\Theta \cup \{ K+1 \}$. The profit for the coalition $\Theta^{+}$ is,
\begin{equation}\label{j}
    \pi_{IO}(\Theta^{+})= \sum\limits_{k \in \Theta}\pi_{k}+\pi_{ISP},
\end{equation}
see \eqref{f} and \eqref{e}. Due to the cancellation of terms involving $t_{k}^{(S)}$ and $t_{k}^{(\beta)}$, which is consistent with the absence of payments by CPs to the ISP in this setting, it follows that
\begin{equation} \label{i}
   \pi_{IO}(\Theta^{+}; \bm{p}, \bm{\beta}, \bm{S}) =\sum\limits_{k \in \Theta}V_{k}(p_k,\beta_k,S_k)-F
\end{equation}
where, for $k=1,2, \ldots ,K,$
\begin{dmath}\label{zz}
   V_{k}(p_k,\beta_k,S_k)=(p_{k}-c-\eta^{(\beta)}\beta_k) \frac{A_k}{p_k^{\varepsilon_k}}\log R_k (\beta_k, S_k)-(c_{k}^{(S)}+\eta^{(S)})S_k-f_k 
\end{dmath}
We interpret $V_k$ as the "operational" profit from a "virtual consortium" of the ISP and CP k. The qualifiers "operational" and "virtual" take note of the absence of the fixed cost F. Also, $V_k$ is the analogue in the coalitional setting of $\pi_k$, in \eqref{e}.
\section{Optimization of Coalitional Business Model}\label{sec3}
We consider the following problem:
\begin{equation}\label{n}
   \max \pi_{IO}(\Theta^{+}; \bm{p}, \bm{\beta}, \bm{S}) \text{ w.r.t } \bm{p}, \bm{\beta}, \bm{S}.
\end{equation}
By setting $\partial\pi_{IO}/\partial p_k=0$ we obtain for the optimal fees for subscribers, $\bm{p^*}$,
\begin{equation}\label{eq:2}
    p_k^*=\frac{\varepsilon_k}{\varepsilon_k-1}(c+\eta^{(\beta)}\beta_k) \text{         }  (k=1,2, \ldots,K)
\end{equation}
The term in parenthesis is the marginal cost for subscribers. We incorporate the above expression in $\pi_{IO}$, and hereafter consider $\pi_{IO}$ and $\{V_k\}$ to be functions of $\bm{\beta}$ and $\bm{S}$ only. That is, from \eqref{i} and \eqref{zz},
\begin{equation} \label{o}
    \pi_{IO}(\Theta^{+}; \bm{\beta}, \bm{S}) =\sum\limits_{k \in \Theta}V_{k}(\beta_k,S_k)-F
\end{equation}
where, for $k=1,2, \ldots, K,$
\begin{equation} \label{k}
    V_{k}(\beta_k,S_k)=a_k(\beta_k)\log R_k (\beta_k, S_k)- (c_{k}^{(S)}+\eta^{(S)})S_k-f_k,
\end{equation}
and,
\begin{equation}
    a_k(\beta_k)=\frac{(\varepsilon_k-1)^{(\varepsilon_k-1)}}{\varepsilon_k^{\varepsilon_k}}\cdot \frac{A_k}{(c+\eta^{(\beta)}\beta_k)^{(\varepsilon_k-1)}}
\end{equation}
Here $a_k (\beta_k)$ represents the profit component from subscriber fees after accounting for the per-subscriber costs for subscriber support and incremental network capacity. The other profit component, $\log R_k$, amplifies the first profit component, and represents the increase in subscribers due to enhancement in QoE from the combination of caching and last mile network capacity increment. The negative terms in \eqref{k} are the virtual coalition's costs for caching and the CP's fixed cost. The problem reduces to,
\begin{equation}
    \max V_k(\beta_k,S_k) \text{   }  (k=1,2, \ldots,K)
\end{equation}
with respect to cache size $S_k$ and capacity increment  $\beta_k$.

\subsection{Optimization with Respect to Cache Size}\label{sec3-1}
From \eqref{k},
\begin{equation}\label{m}
    \frac{\partial V_k(\beta_k,S_k)}{\partial S_k}= \frac{a_k(\beta_k)(\delta r+\beta_k)}{(\log\Sigma_{k})R_k}\cdot\frac{1}{(S_k+1)}-(c_k^{(S)}+\eta^{(S)})
\end{equation}
Since the right hand side decreases with increasing $S_k$, $V_k$ is concave in $S_k>0$. Moreover, if $\partial V_k  (\beta_k,0)/\partial S_k  \leq 0$, then the maximum of $V_k$ with respect to $S_k$ is achieved when $S_k = 0$. We have
\begin{prop}
A necessary and sufficient condition for $V_k (\beta_k, S_k)$ to reach its maximum value at a unique positive value of $S_k$ is
\begin{equation}\label{l}
    (c_k^{(S)}+\eta^{(S)})< \frac{a_k(\beta_k)(\delta r+\beta_k)}{(\log\Sigma_{k})}
\end{equation}
If the inequality holds, then the unique optimum value of $S_k$  is obtained from the first order condition $\partial V_k/\partial S_k  = 0$. If the inequality in \eqref{l} is violated, then the optimal $S_k=0$.
\end{prop}
The proof follows from the concavity of $V_k (\beta_k, S_k)$ with respect to $S_k$; \eqref{l} expresses  $\partial V_k  (\beta_k,0)/\partial S_k > 0$. \par
The interpretation of \eqref{l} is that the left hand quantity is the unit cost of the cache to the ISP and CP k, and the right hand quantity is the marginal profit from caching at $S_k =0$.
\subsection{Optimization of Incremental Network Capacity} \label{sec3-2}
The reader may verify, from \eqref{k}, that
\begin{equation}\label{z}
    \frac{\partial V_k(\beta_k,S_k)}{\partial \beta_k} = \psi_k(\beta_k,S_k)a_k(\beta_k)\log R_k(\beta_k,S_k)
\end{equation}
where,
\begin{equation}
    \psi_k(\beta_k,S_k)=\Bigg[ \frac{h_k}{R_k(\beta_k,S_k)\log R_k(\beta_k,S_k)} -\frac{(\varepsilon_k-1)\eta^{({\beta})}}{(c+\eta^{(\beta)}\beta_k)} \Bigg]
\end{equation}
Note from \eqref{z} that the sign of $\psi_k(\beta_k,S_k)$ is the sign of $\frac{\partial V_k(\beta_k,S_k)}{\partial \beta_k} $ since the other factors are positive. Also, as $\beta_k \to \infty, \psi_k(\beta_k,S_k) \to 0 $.\par
 If $\psi_k(0,S_k)>0$, then, since $\psi_k$ is continuous in $\beta_k$, there will be at least one positive solution to $\partial V_k(\beta_k,S_k)/\partial \beta_k =0 $. Let the smallest positive solution be $\beta_k^*$, where $V_k(\beta_k,S_k)$ will be maximized (possibly locally). Hence, we have
\begin{prop}
 If $\psi_k(0,S_k)>0$, i.e,
 \begin{equation}\label{eq:4}
     \frac{\eta^{({\beta})}}{c}<\frac{1}{(\varepsilon_k-1)}\frac{h_k}{(\delta r h_k+1)\log (\delta r h_k+1))}
 \end{equation}
then there exists at least one stationary maximum point of $V_k (\beta_k, S_k)$ at $\beta_k^* > 0$.
\end{prop}
Note that the left hand side is the ratio of  $\eta^{(\beta)}$  , the unit cost of the incremental network capacity over the last mile, and c, the per-subscriber cost.

\subsection{Characteristic Function of the Game}\label{sec3-3}
For any coalition $\Theta$, the characteristic function value $v(\Theta)$ is the profit that the coalition generates. If the coalition does not include the ISP (indexed K+1), the characteristic function value is null, as is also the case for the singleton \{K+1\}. We now consider a coalition of CPs and ISP, i.e., $\Theta \subseteq \mathbb{K},$ and $\Theta^+ = \Theta \cup \{K+1\}$.\par
Let the maximization of $V_k$ with respect to $\beta_k$ and $S_k$ by the above procedures yield,
\begin{equation}\label{eq:sp1}
    v_k=\max V_k(\beta_k,S_k) \text{ , }  (k=1,2, \ldots,K)
\end{equation}
From \eqref{o},
\begin{equation}
    \max\limits_{\bm{p}, \bm{\beta}, \bm{S}} \pi_{IO}(\Theta^{+}; \bm{p}, \bm{\beta}, \bm{S}) = \sum\limits_{k \in \Theta}v_k-F
\end{equation}
Hence, the the characteristic function,
\begin{equation}\label{p}
    v(\Theta^+) = \sum\limits_{k \in \Theta}v_k-F
\end{equation}
Note that if  $\Theta \cup \{k\}= \varnothing$, then
$ v(\Theta^+ \cup \{k\})- v(\Theta^+)=v_k>0$,
so that the characteristic function of the game is super-additive.
\section{Coalitional Bargaining} \label{sec4}
\subsection{Review of Related Models}
Our interest is in the formation of coalitions, from the set of players composed of an ISP and multiple CPs, within the framework of multi-player, non-cooperative game theory. As in two-player bargaining [23,24], the multi-player bargaining game may be approached either via an axiomatic model, such as the Nash bargaining model, or a strategic model [12],[25,26]. In these games bargaining proceeds in rounds with time discounting, i.e., payoffs are discounted by a factor $\delta(\delta<1)$ with each additional round. In general, a player makes a proposal, and the other players in the proposed coalition either accept or reject the proposal. In the latter case, the game proceeds to the next round, with a player making another proposal. The order in which players make proposals has significant consequences; for instance, the advantage of the first proposer is well-known [24]. Hence models may be classified into "fixed order" games, where the protocol determines the order in which players make proposals, and "random order" games in which proposers are chosen uniformly at random in each round. All the models discussed below use the Stationary Subgame Perfect Equilibrium (SSPE) as a solution concept. (Subgame Perfect Equilibrium (SPE) is a Nash equilibrium of the original game which is also a Nash equilibrium in every subgame of the game. A solution of the game is called SSPE if it is a SPE with the property that at any round the strategy of every player depends only on the set of all players active in that round.) \par
Chatterjee et. al. [25] consider fixed order protocols in which the first rejector becomes the new proposer, and focus on bargaining games with equilibrium that is efficient asymptotically, i.e. efficient as $\delta \to 1$. They show that two properties of asymptotically efficient equilibrium, namely, Grand Coalition formation and no delay, may be violated by the stationary equilibrium if the "egalitarian" payoff vector (in which all payoffs are equal) is not in the core of the game.\par
Compte and Jehiel [26] investigate a bargaining model in which only one coalition can form (this is not assumed by Okada [12]). Therefore, in super-additive games, players bargain under the threat that a smaller team can form. In this model it is assumed that players adopt mixed strategies, as opposed to pure strategies assumed in [12] and [25]. The main result in [26] is that if an asymptotically efficient stationary equilibrium exists, then the payoff vector in the equilibrium is identical to the one obtained by maximizing the Nash product, which is the product of the players' payoffs.
\subsection{Okada's Bargaining Model and Results}
In the bargaining procedure of \cite{b18}, at every round $t, t = 1,2,\ldots,$ one player, which may be either a CP or the ISP, is selected as a proposer with equal probability among all players still active in bargaining. Let $\mathbb{K}^t$  be the set of active players at round t, where $\mathbb{K}^1$ = $\mathbb{K}^+$. The selected player k proposes (i) a coalition set $\Theta^+ $ with $k \in \Theta^+ \subseteq \mathbb{K}^t$ and $v(\Theta^+ ) > 0,$ and (ii) a payoff vector $x(\Theta^+)$. All other players in $\Theta^+$ either accept or reject the proposal sequentially. If all the other players in the coalition accept the proposal, then it is agreed upon, and all the remaining players outside $\Theta^+$ can continue negotiations at the next round. Otherwise, negotiations go on to next round, and a new proposer is chosen uniformly at random. The process continues until there is no possibility of a coalition with positive value.\par
The payoffs of players are determined as follows. When a proposal $(\Theta^+,x(\Theta^+))$ is agreed upon at round t, the payoff of every member k in $\Theta^+$ is $\delta^{t-1} x(\Theta^+)$, where $\delta$ is the discount factor. For players who fail to join any coalition, their payoffs are assumed to be zero. A perfect information model is assumed, i.e., each player has perfect information about the history of the game whenever the player makes a decision. \par
In addition to the Stationary Subgame Perfect Equilibrium (SSPE), we need the concepts of (i) limit SSPE, and (ii) limit subgame efficient SSPE. Let the game defined above be denoted by $G (\mathbb{K}^+, \delta)$, and subgames by $G (\Theta^+, \delta)$. A limit SSPE is a limit point of SSPEs of $G (\mathbb{K}^+, \delta)$ as $\delta$ goes to 1. A SSPE of $G (\mathbb{K}^+, \delta)$ is called subgame efficient if for every subgame $G (\Theta^+, \delta)$, every player proposes the full coalition $\Theta^+$ in the SSPE. A limit subgame efficient SSPE is defined to be a limit point of subgame efficient SSPEs of $G (\mathbb{K}^+, \delta)$ as $\delta$ goes to 1. The notion of subgame efficiency is a strong one as it requires that in all possible rounds of negotiations every player proposes the full coalition of the active players.
\begin{theorem}[Okada]
There exists a limit subgame efficient SSPE of the game (up to the response rule of the equilibrium path) if and only if the game satisfies
    \begin{equation*}
        \frac{v(\Psi^+)}{|\Psi^+|} \geq \frac{v(\theta^+)}{|\theta^+|} \text{ for all } \theta^+  \subseteq \Psi^+ \subseteq \mathbb{K^+}
    \end{equation*}
The expected equilibrium payoff vector is egalitarian, i.e., $w(\Theta^+ ) =( \frac{v(\Theta^+)}{|\Theta^+|},\frac{v(\Theta^+)}{|\Theta^+|} ,\ldots, \frac{v(\Theta^+)}{|\Theta^+|}) )$ in every subgame  $G (\Theta^+)$.
\end{theorem}
The result shows that the subgame efficient SSPE of the bargaining game $G (\mathbb{K}^+, \delta)$ exists for any sufficiently large $\delta$ if and only if the per capita profit $PC(\Theta^+)=v(\Theta^+)/|\Theta^+|$ increases with coalition size. If the condition in the Theorem holds, then the limit efficient SSPE will also hold for every fixed order protocol in the bargaining model in Chatterjee et. al. \cite{b17}, and hence, as noted by Okada, the result is robust with respect to changes in the rule governing the selection of proposers.
\section{Grand Coalition Formation and Properties}\label{sec5}
We obtain the necessary and sufficient condition for the formation of the Grand Coalition as stated in Theorem 1. We prove various coalition properties, including for the scenario in which the coalition is dynamically evolving. We also compare the payoffs to the coalition partners in our non-cooperative game-theoretic setting to Shapley Value payoffs, which are derived from a fair profit-allocation scheme in cooperative games \cite{b21}.
\subsection{Grand Coalition Formation With Random Proposers}\label{sec5-1}
For any coalition $\Theta^+$, the per-capita profit, 
\begin{equation}
    PC(\Theta^+) = \frac{1}{(|\Theta^+|)}v(\Theta^+)=\frac{1}{(|\Theta|+1)}\Big[\sum\limits_{k\in \Theta}v_k-F \Big]
\end{equation}
Without loss of generality, we index the players such that the following holds,
\begin{equation} \label{t}
    v_1\leq v_2 \leq v_3 \leq \ldots \leq v_K
\end{equation}
To establish the key condition in Theorem 1, namely, the per-capita profit increases with increasing coalition size, consider any two coalitions, $\Theta$ and $\Psi$, such that $\Theta \subset \Psi \subseteq \mathbb{K} . \text{ Let } \Theta_c= \Psi \setminus\Theta$
\begin{prop}
A necessary and sufficient condition for $ PC(\Psi^+) \geq PC(\Theta^+) $ is
\begin{equation} \label{u}
    v_1 \geq \frac{1}{(K+1)}\Big[\sum\limits_{k =1}^{K}v_k-F\Big]
\end{equation}
\end{prop}
\textit{Proof:}
\begin{align*}
    & PC(\Theta^+) - PC(\Psi^+) = \\
    & \frac{1}{(|\Theta|+|\Theta_c|+1)}\Big[\sum\limits_{k \in \Theta}v_k+\sum\limits_{k \in \Theta_c}v_k-F \Big]-\frac{ \Big[\sum\limits_{k \in \Theta}v_k-F \Big] }{(|\Theta|+1)}   \\
    &= \frac{|\Theta_c|}{(|\Theta|+|\Theta_c|+1)}\Bigg[\frac{1}{|\Theta_c|}\sum\limits_{k \in \Theta_c}v_k - \frac{1}{(|\Theta|+1)}\Big[\sum\limits_{k \in \Theta}v_k -F\Big] \Bigg] \\
    &\geq   \frac{|\Theta_c|}{(|\Theta|+|\Theta_c|+1)} \Bigg[\frac{1}{|\Theta_c|}\sum\limits_{k =1}^{|\Theta_c|}v_k - \frac{1}{(|\Theta|+1)}\Big[\sum\limits_{\substack{k ={}\\ \tiny{ K-|\Theta|+1}}}^{K}v_k -F\Big] \Bigg] \numberthis \label{q}\\
    &\geq   \frac{|\Theta_c|}{(|\Theta|+|\Theta_c|+1)} \Bigg[v_1 - \frac{1}{(|\Theta|+1)}\Big[\sum\limits_{k =K-|\Theta|+1}^{K}v_k -F\Big] \Bigg] \\
    &\geq   \frac{|\Theta_c|}{(|\Theta|+|\Theta_c|+1)} \frac{K}{(|\Theta|+1)}\Bigg[v_1 - \frac{1}{K}\Big[\sum\limits_{k =2}^{K}v_k -F\Big] \Bigg] \numberthis \label{r}\\
    &=\frac{|\Theta_c|}{(|\Theta|+|\Theta_c|+1)} \frac{(K+1)}{(|\Theta|+1|)}\Bigg[v_1 - \frac{1}{K+1}\Big[\sum\limits_{k =1}^{K}v_k -F\Big] \Bigg] \numberthis \label{s}
\end{align*}
The proof of \eqref{q} follows from the ordering in \eqref{t}, and \eqref{r} follows from
\begin{equation}
    \sum\limits_{k =2}^{K}v_k-Kv_1 \geq \sum\limits_{k =K-|\Theta|+1}^{K}v_k-(|\Theta|+1)v_1
\end{equation}
Hence, from \eqref{s}, the condition in \eqref{u} is a sufficient condition for $PC(\Psi^+) \geq PC(\Theta^+)$. 
Necessity follows from the case $\Theta = \mathbb{K}\setminus \{1\} \text{ and } \Psi = \mathbb{K}$. \par
We may now state \eqref{u} without reference to the particular ordering in \eqref{t}. Let the minimum and average characteristic function values be denoted thus,
\begin{equation}
    v_{min}(\mathbb{K})=\min\limits_{1\leq k \leq K}v_k  \text{ and } v_{avg}(\mathbb{K})=\frac{1}{K}\sum\limits_{k =1}^{K}v_k
\end{equation}
\begin{cor}
The following conditions are equivalent to the condition \eqref{u},
\begin{equation} \label{y}
    v_{min}(\mathbb{K}) \geq \frac{1}{(K+1)}v(\mathbb{K}^+)
\end{equation}
\begin{equation} \label{w}
   \frac{1}{(K+1)} [v_{avg}(\mathbb{K})+F] \geq v_{avg}(\mathbb{K})-v_{min}(\mathbb{K})
\end{equation}
\end{cor}
\par To recapitulate, if and only if either \eqref{y} or \eqref{w} holds then the Grand Coalition forms, and in this case the payoff vector is egalitarian, i.e., all players get an equal share. It is easily verified that the egalitarian payoff vector is in the core of the game.  We denote the payoff to player k in the Grand Coalition by $w_k (k=1,2,\ldots, K+1)$, so that, for all k,
\begin{equation}\label{d1}
    w_k=\frac{1}{K+1}v(\mathbb{K}^+)=\frac{1}{K+1}\Big[\sum\limits_{k=1}^{K}v_k-F\Big]
\end{equation}
An interpretation of the above conditions is that subsidies from any coalition of the bigger contributors to any coalition of smaller contributors may exist, but these subsidies are bounded. The role of the common fixed cost, F, is noteworthy. It plays an essential role in justifying the coalition. Also note that on account of the factor, $1/(K+1)$, on the left hand side of \eqref{w}, assuming that all other quantities are held fixed, the formation of larger coalitions becomes increasingly hard.
\subsection{Grand Coalition Formation Procedure}
To recapitulate the preceding results on the Grand Coalition formation, the following two steps are undertaken sequentially for any candidate set of CPs and an ISP.
\begin{itemize}
\item \textit{Design Optimization.} As described in Sec. \ref{sec3}, the optimum contribution $v_k$ of CP k to the characteristic function is computed for each CP, see \eqref{eq:sp1}. The nearly separable structure of the coalition's profit function, see \eqref{o}, allows its optimization to be decomposed into K simpler design optimizations, each of which is specific to a CP and the ISP. 
\item \textit{Coalition Admission Control.}  This step ensures that either condition \eqref{y} or \eqref{w} is satisfied by all admitted CPs. This may require denying admission to the coalition to certain players, e.g., the smallest contributors among the CPs. Negotiations on coalition formation proceed with the players restricted to the admitted set. Thus, the Grand Coalition that forms is composed exclusively of the admitted players.
\end{itemize}
Of course, this raises various related questions, some of which are addressed below and others that are outside the scope of this paper. For instance, might it be possible for a group of CPs to band together to form a new entity which will satisfy the admission control?
\subsection{Settlements}
We make explicit the settlements that must take place for the Grand Coalition's foundational goal of equal payoffs to be satisfied. The settlements that we obtain follow from book-keeping of the revenue and costs of each player in the coalitional business model. Since the only revenue is from CPs' subscribers, the settlements will involve payments from CPs to the ISP. Let $T_k$ denote the payment from CP k to the ISP as settlement. Then,
\begin{equation} \label{b1}
    T_k=(p_k-c)n_k-c_k^{(S)}S_k-f_k-w_k
\end{equation}
For the ISP, it may be verified that it's payoff in the coalition,
\begin{equation}
     w_{K+1} = \sum\limits_{k=1}^{K}T_k-\eta^{(\beta)}\sum\limits_{k=1}^{K}\beta_kn_k-\eta^{(S)}\sum\limits_{k=1}^{K}S_k-F
\end{equation}
In the context of the non-coalitional business model in \ref{sec2-3}, observe that  
\begin{equation}\label{zy}
    T_k=\{t_k^{(S)}S_k+t_k^{(\beta)}\beta_kn_k \} \text{   } (k=1,2,\ldots K)
\end{equation}
That is, the settlement payments in the coalition may be interpreted to be payments by CPs to the ISP for cache and network capacity. Note, importantly, that \eqref{zy} does not uniquely determine the unit prices $t_k^{(S)}$ and $t_k^{(\beta)}$.
\subsection{Grand Coalition Sustainability in Online Environment}
We consider the sustainability of the Grand Coalition in a dynamic setting where CPs may apply to dropout and other CPs may apply to join.
\begin{prop}
If either \eqref{y} or \eqref{w} in the Corollary to Proposition 3 holds, then, irrespective of the number and identities of CPs which may dropout of the coalition, the condition will continue to hold for the remaining members.\\
\end{prop}
\textit{Proof:} Follows immediately from the fact that \eqref{y} ensures that the per capita profit of any coalition increases with coalition size. This property remains intact with the remaining members. \par
On the other hand, if a new CP applies to join, the satisfaction of \eqref{y} is not guaranteed after its inclusion, and therefore needs to be checked as part of the Coalition Admission Control discussed in Sec. \ref{sec5-1}.
\subsection{Net Neutrality}
In our model, caching may indeed be supported even when Net Neutrality regulations apply and consequently additional network capacity of $\beta_k$ to individual users of CP k is not allowed.  It is easy to see by examining \eqref{l} for the existence of positive cache sizes in the optimized design, that this condition may be satisfied even with $\beta_k = 0$. Numerical results in Sec. \ref{sec6} will bear this out.\par
Prior results \cite{b7} have shown that when an ISP and a CP interact according to the Stackelberg leader-follower model, caching is not supported when Net Neutrality regulations apply and $\beta_k = 0$. Our results show that this negative result is an artifact of the Stackelberg model, and not intrinsic to Net Neutrality regulations.
\subsection{Shapley Value}
Shapley Value is a concept for "fair" profit-sharing in cooperative game theory \cite{b21}. In our setting games are non-cooperative. Nonetheless, it is interesting to compare and contrast payoffs.\par
The Shapley Value payoff to a player is determined by the average increment in profit that is due to its inclusion in coalitions, with the average computed over all coalitions
\begin{prop}
Let $x_k$ denote the Shapley Value payoff to player k.
\begin{equation} \label{e1}
    x_k=\frac{1}{2}v_k-\frac{1}{K(K+1)}F \text{ , } (k=1,2\ldots,K)
\end{equation}
\begin{equation} \label{f1}
    x_{K+1}=\frac{1}{2}\sum\limits_{k=1}^{K}v_k-\frac{K}{K+1}F
\end{equation}
\end{prop}
We omit the proof since it is based on straightforward counting.\par
These quantities may be compared to the payoff $w_k=[\sum\limits_{k=1}^K v_k   -F]/(K+1)$ to all players in the coalition. Clearly, in the Shapley Value the ISP bears a substantially larger burden of the fixed cost F. On the other hand, half of the value generated by the virtual coalition (see Sec. \ref{sec3}) composed of CP k and the ISP goes towards the CP's payoff, and the other half to the ISP. It is interesting how different foundations of business relationships give rise to different payoffs even when they share the common goal of fair and equitable allocations. Numerical results in Sec. \ref{sec6} will illustrate the allocations.

\section{Numerical Results}\label{sec6}
We report on numerical studies on characteristics of the Grand Coalition, the egalitarian payoff relative to payoffs in the Shapley Value, and the effects of Net Neutrality. Table \ref{tab1} gives the base model parameters, which are used throughout this section except when explicitly stated otherwise. In the base case, there are five homogeneous CPs.
\begin{table}
\begin{center}
    \begin{tabular}{|c|c|c|c|c|c|c|c|c|c|c|}
    \hline
    $A_k$ &  $\sum_k$ &$c$  & $n_{\beta}$ &  $n_{S} $ &    $c_{S}$ &  $r_1$ &  $r_2$ & $K$ &  $F$ & $f_k$  \\
    \hline
    $2 \times 10^5$ &  $10^5$ &  1 & 0.1 & 0.5 & 0.5 & 4 & 1 & 5 & $10^5$ & $10^4$ \\
    \hline
    \end{tabular}
    \caption{List of parameters}
    \label{tab1}
\end{center}
\end{table}
\subsection{Design for Profit Maximization}
\begin{figure}[htp]
  \centering
  \subfigure[$v(\mathbb{K}^+)$ vs $\varepsilon$  at $\eta^{(S)}=0.5$ \label{niCPs}]{\includegraphics[height=4.5cm, width=7cm]{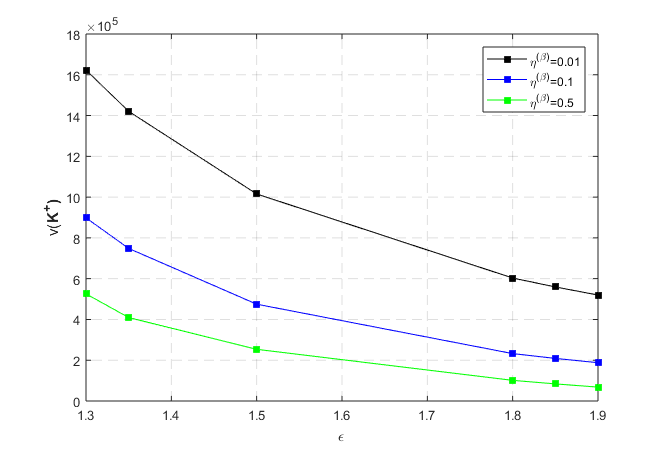}}
  \subfigure[$v(\mathbb{K}^+)$ vs $\varepsilon$  at $\eta^{({\beta})}=0.1$ \label{niCPs_1}]{\includegraphics[height=4.5cm, width=7cm]{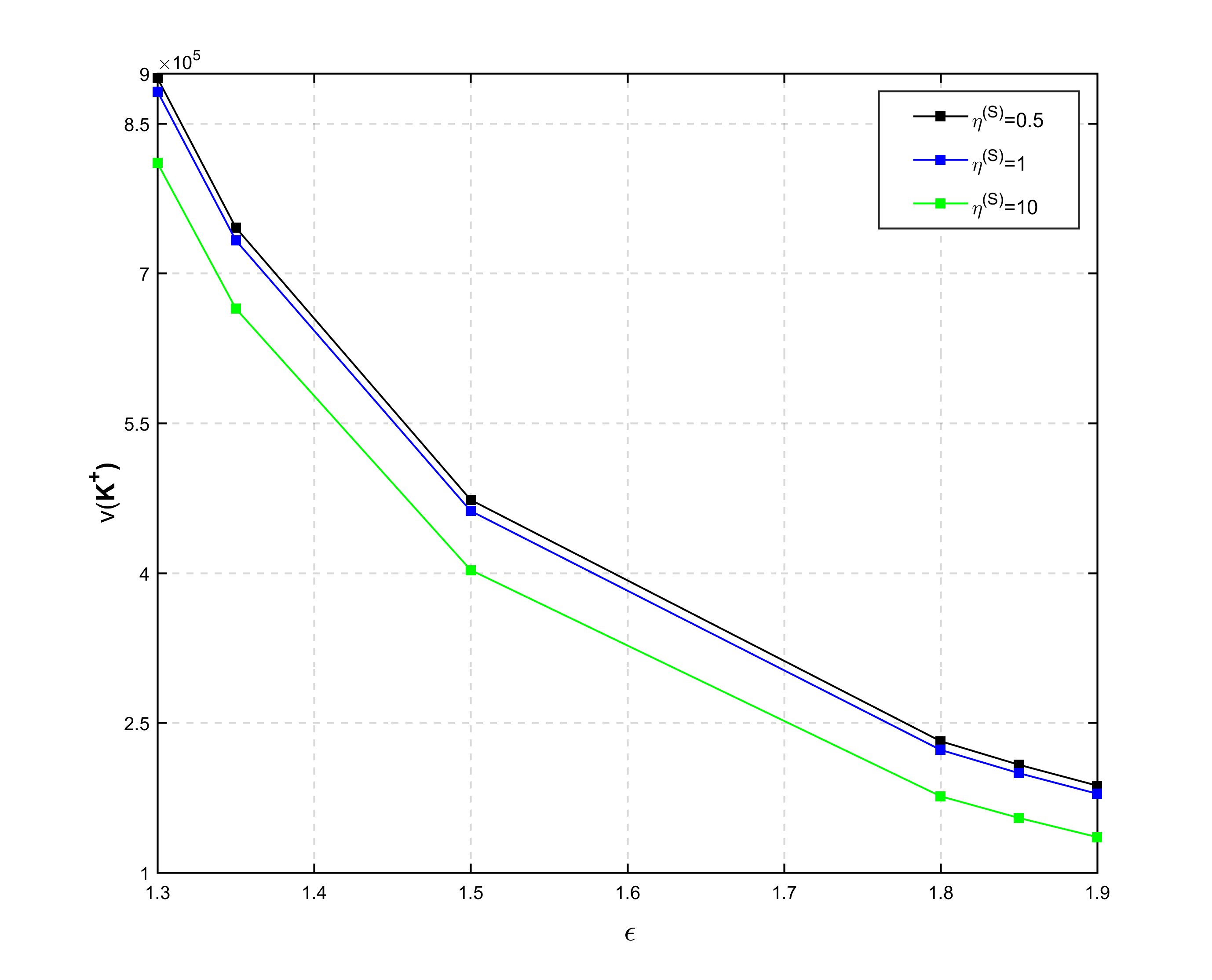}} 
  \caption{Graph showing optimized coalition profit as a function of $\varepsilon$}
\end{figure}
\begin{figure}[htp]
  \centering
  \subfigure[Dependence of $V_k$ on $S_k$ at $\epsilon=1.5$ and $\eta^{(\beta)}=0.1$\label{cvx2}]{\includegraphics[height=4.5cm, width=7cm]{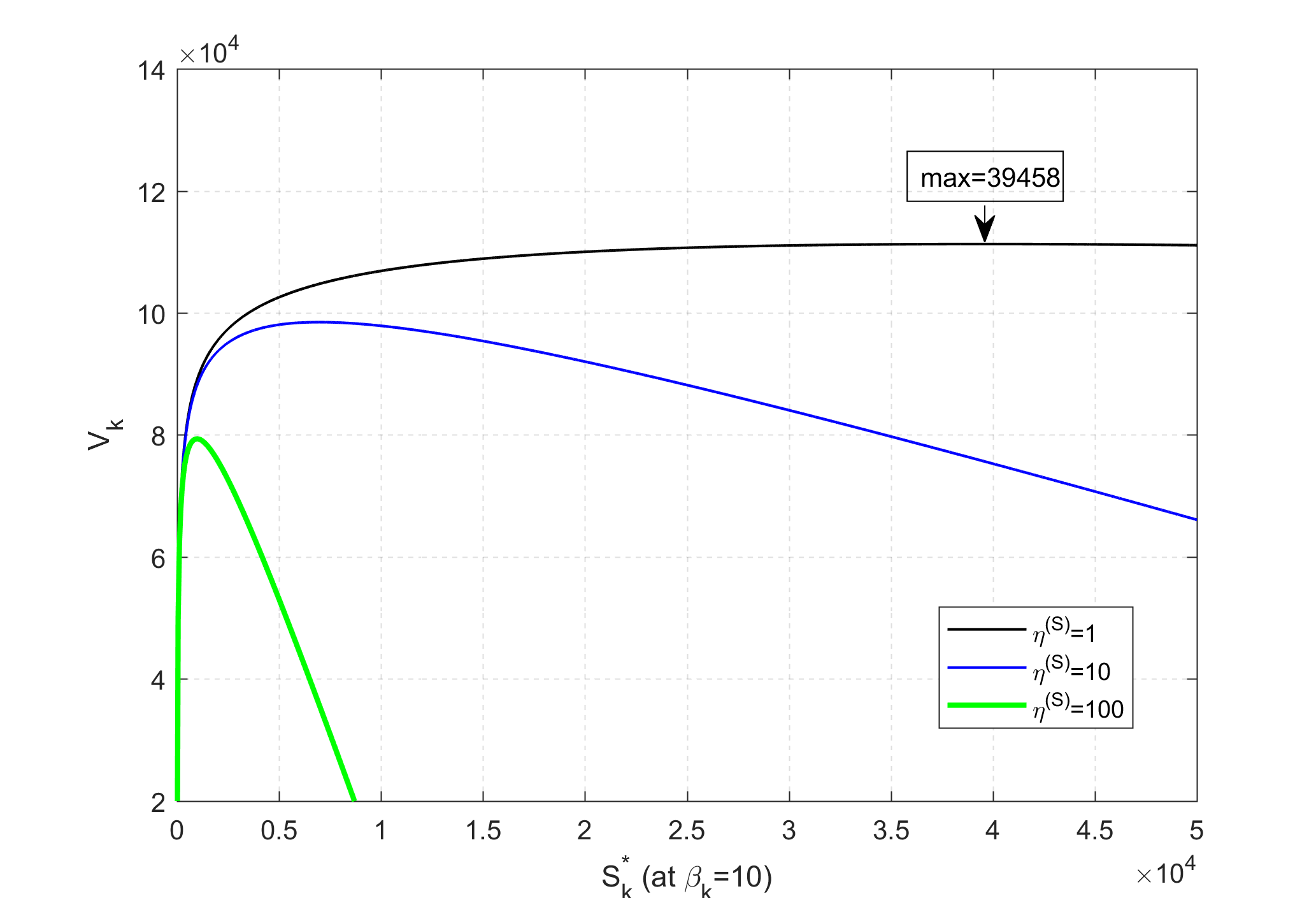}}
  \subfigure[Dependence of $V_k$ on $\beta_k$ at $\epsilon=1.5$ and $\eta^{(S)}=1$ \label{cvx1}]{\includegraphics[height=4.5cm, width=7cm]{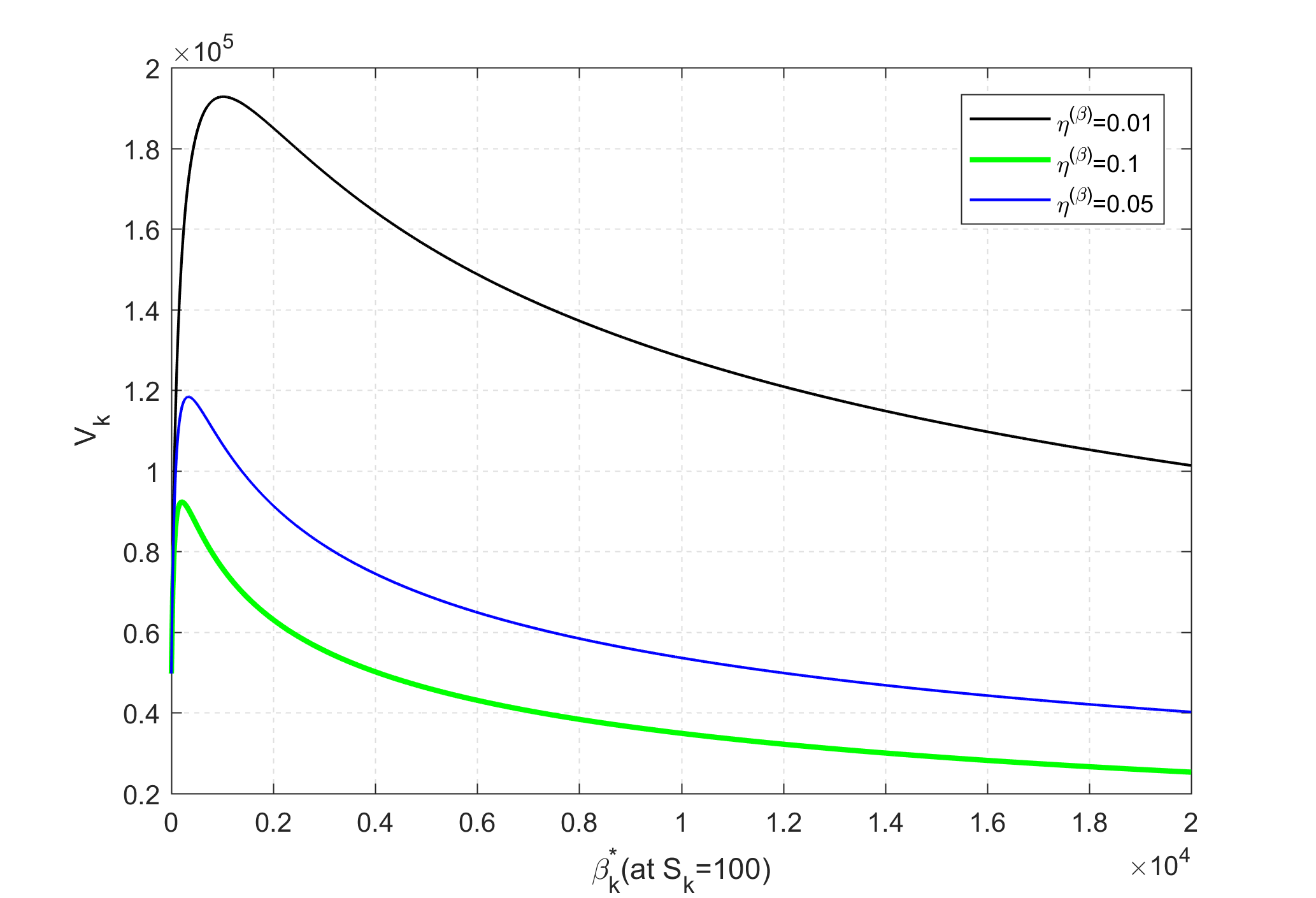}}
  \caption{Graph indicating existence of optimum $\beta$ and $S$}
\end{figure}
\begin{figure}[htp]
  \centering
  \includegraphics[height=4.5cm, width=7cm]{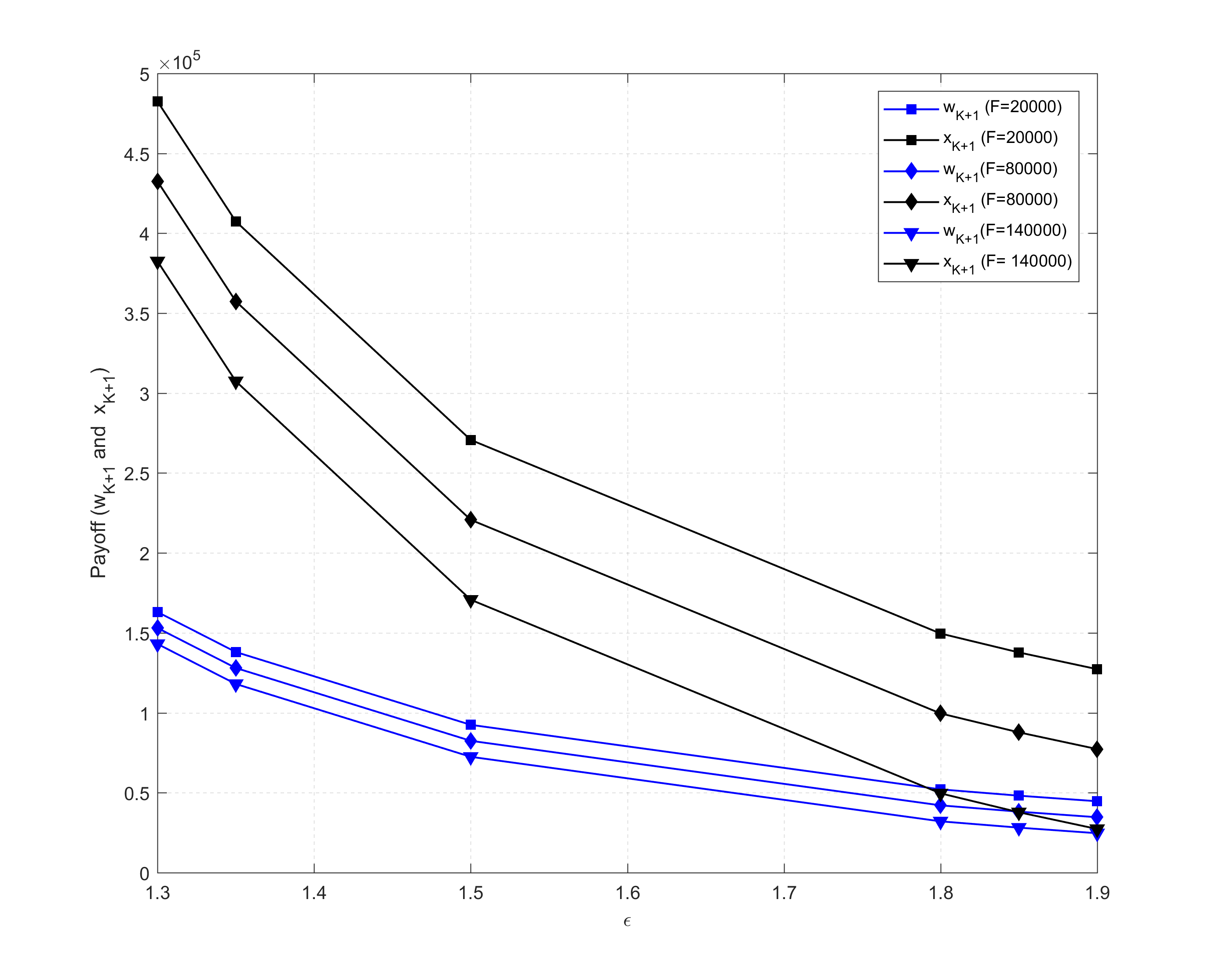}
  \caption{ISP's Nash Bargaining and Shapley Value payoffs at different fixed costs}\label{sh}
\end{figure}
\begin{table}
\begin{center}
\begin{tabular}{|c|c|c|c|c|}
     \hline
     Parameter &$\eta^{(S)}=1$ &$\eta^{(S)}=100$ &$\eta^{(S)}=10^{4}$ &$\eta^{(S)}=10^{5}$ \\
     \hline
     $p $&7.32 &9.3&22.59&3\\
     \hline
     $h$ &0.710&0.388&0.079&0\\
    \hline
    $S$& 3699.7 &86.8 & 1.5 & 0 \\
    \hline
    $\beta$ &14.4 &	21 & 65.3 &0\\
    \hline
    $v(\mathbb{K}^+) \cdot 10^4$ &	12.241 & 9.336 & 3.723 & 0 \\
    \hline
\end{tabular}
\caption{Coalition profit and design parameters after optimization in NNN case}
\label{tableNNN}
\end{center}
\end{table}

\begin{table}
\begin{center}
\begin{tabular}{|c|c|c|c|c|}
     \hline
    Parameter & $\eta^{(S)}=1$ & $\eta^{(S)}=100$ &	$\eta^{(S)}=10^{4}$&$\eta^{(S)}=10^{5}$ \\
     \hline
     $p$	&3	&3	&3 &3\\
     \hline
     $h$	&0.7249	&0.3935	& 0.0501 & 0 \\
    \hline
    $S$	& 4211		& 90.7		& 0.08&0 \\
    \hline
    $\beta$ &	$0$	& $0$&	$0$ & $0$\\
    \hline
    $v(\mathbb{K}^+)\cdot 10^3$ &	82.612 & 50.786 & 2.970 & 0\\
    \hline
\end{tabular}
\caption{Coalition profit and design parameters after optimization in NN case}
\label{tableNN}
\end{center}
\end{table}
\vspace{1mm}
Our results here are from the optimization of the coalitional business model treated in Sec. \ref{sec3}. The dependence of the maximum profit on the price elasticity $\varepsilon_k$ is not \textit{a priori} obvious. Note that increasing $\varepsilon_k$ lowers the demand, see (5); however the factor $\varepsilon_k$/($\varepsilon_k$ -1) in (18) suggests that the price increases and demand decreases. Moreover, the dependence of the optimum price on the amount of incremental network capacity $\beta_k$ is a further complication.\par
Figures \ref{niCPs} and \ref{niCPs_1} show the afore-mentioned dependence for various values of the unit costs for cache and network capacity, $\eta^{(S)}$ and $\eta^{(\beta)}$, respectively. The figures show that the optimum profit $v(\mathbb{K}^{+})$  decreases with increasing elasticity; the optimum values of the price $p_k^*$, cache  $S_k^*$  and incremental network capacity $\beta_k^*$  also decrease. Also, $v(\mathbb{K}^{+})$ is more sensitive to changes in $\eta^{(\beta)}$ than $\eta^{(S)}$. \par
Fig. \ref{cvx2} shows the behavior of the profit $V_k$  as the cache size $S_k$ is varied for fixed price elasticity of demand of 1.5 and incremental network capacity of 10, for various values of the unit cost of cache. The plots corroborate the concavity of $V_k$ as a function of $S_k$, which is proven in Sec. \ref{sec3-1}. Similar to Fig. \ref{cvx2}, Fig. \ref{cvx1} shows the behavior of $V_k$  as the incremental network capacity $\beta_k$ is varied. Concave behavior is notably missing; however, note the existence of a stationary maximum point, the subject of Proposition 2.
\subsection{Shapley Value}
Fig. \ref{sh} compares the payoff to the ISP in the coalition to its Shapley Value payoff. As noted earlier, in the Shapley Value the ISP bears almost all of the fixed cost F and half of each CP's fixed cost. Although the coalitional bargaining game is in the framework of non-cooperative games, the coalition, once formed, operates as a joint operation, i.e., the ISP and CPs behave as a single business entity, and all the fixed costs for setting up and operating the infrastructure are shared equally.
\subsection{Net Neutrality}
We explore the impact of Net Neutrality (NN) regulations on the formation of coalitions. To recapitulate, our assumption is that with NN the incremental network capacity is zero. In the example below we contrast NN with with Non-Net Neutrality (NNN), the base case in this paper. In this example the price elasticity of demand $\varepsilon = 1.3$, the parameters in Table \ref{tab1} apply except the fixed cost of the fifth CP, $f_5$ is increased to $7.0 \times 10^4$, which has the effect of reducing this CP's contribution $v_5$. (Recall that this contribution is obtained after design optimization.) \\
\textbf{Case 1 : NNN}\\
$v(\mathbb{K}^+)=8.3843\times 10^5$\\
$v_{5}=1.3997\times 10^5 > 1.3974\times 10^5=\frac{v(\mathbb{K}^+)}{6}$\\
Hence in this case with NNN, the Coalition Admission Control is satisfied by the set of five CPs, and the Grand Coalition forms.\\ 
\textbf{Case 2 : NN}\\
$v(\mathbb{K}^+)=3.4939\times 10^5$\\
$v_{5}=4.2033\times 10^4 < 5.8232 \times 10^4=\frac{v(\mathbb{K}^+)}{6}$\\
Hence in this case with NN, the Grand Coalition does not form with all five CPs.. 
\par Next we provide some insight into the contrasting effects of NN and NNN on optimized system parameters, such as price, cache size and incremental network capacity in the special case of a single CP, price elasticity = 1.5, and zero fixed costs, i.e., $F = f_1 = 0$. The results for various values of the unit cost of cache, $\eta^{(S)}$, are shown in Tables \ref{tableNNN} and \ref{tableNN}. It can be seen that NNN leads to higher prices, QoE and profits. Also, importantly, in the NNN case cache $S$ and incremental network capacity $\beta$ behave as substitute resources \cite{b8}. That is, as the unit cost of cache $\eta^{(S)}$ increases, the cache installed decreases while the incremental network capacity increases. Also note that at very high unit cost of cache, for both NN and NNN, the installed cache is zero. The reason is that (24) is no longer satisfied.
\section{Conclusion} 
We conclude by noting that the case made here for cost-sharing consortiums applies to settings beyond ISP – CP interworking. Economic theory informs us that whenever large fixed costs are incurred by single producers, it makes for "natural monopolies" since a single producer can then meet market demand at lower cost than multiple producers \cite{b14}, \cite{b20}. However, this cost advantage comes at a high price for society. The collateral downside includes, as is well-known, high barriers to entry to new entrants, and reduced investments in innovations and R\&D \cite{b28}, \cite{b29}. Thus, the case made here for consortiums which share costs should be considered broadly, for instance, in newly developing areas, such as the sharing economy and smart grids.


\begin{thebibliography}{00}
\bibitem{b1}	TestInternetSpeed.org, "Half of All Internet Traffic Goes to Netflix and YouTube", July 9, 2018 [Online], Available: http://testinternetspeed.org/blog/half-of-all-internet-traffic-goes-to-netflix-and-youtube/

\bibitem{b2} Statista, "Statistics and Facts on Netflix", [Online]. Available: https://www.statista.com/topics/842/netflix/

\bibitem{b3}	T. Leighton, "Improving Performance on the Internet", Communications of the ACM, February 2009, Vol. 52 No. 2, Pages 44-51

\bibitem{b4}	K. Russell, "What the Netflix-Comcast Deal Really Means in Plain English", Business Insider, Feb 23, 2014 [Online]. Available: http://www.businessinsider.com/netflix-comcast-deal-explained-2014-2

\bibitem{b5}	Netflix, "Netflix Open Connect" [Online]. Available: https://openconnect.netflix.com/en/

\bibitem{b6}	S. Cheshire, "It's the Latency, Stupid" [Online]. Available: http://www.stuartcheshire.org/rants/latency.html

\bibitem{b7} D.Mitra, Q. Wang and A.Hong, "Emerging Internet Content and Service Providers' Relationships: Models and Analyses of Engineering, Business and Policy Impact", Proc. INFOCOM 2017- IEEE Conference on Computer Communications, Atlanta, GA.

\bibitem{b8} H.R.Varian, "Intermediate Microeconomics with Calculus: A Modern Approach", W.W. Norton and Co., New York NY, 2014.

\bibitem{b9}	S. Crawford, "Captive Audience: The Telecom Industry and Monopoly Power in the Gilded Age", Yale University Press, 2013

\bibitem{b14} R. Braeutigam "Optimal Policies for Natural Monopolies", chapter 23 in "Handbook for Industrial Organizations", vol. 2, ed. R. Schmalensee and R.D. Willig, Elsevier North Holland, Amsterdam, pp 1289-1346, 1989.

\bibitem{b20} J.C. Panzar, "Technological Determinants of Firm and Industry Structure", chapter 1 in Handbook for Industrial Organizations", vol. 1, ed. R. Schmalensee and R.D. Willig, Elsevier North Holland, Amsterdam, pp 3-59, 1989.

\bibitem{b18} A. Okada, "A Noncooperative Coalitional Bargaining Game with Random Proposers", Games and Economic behavior, 16(1):97-108, 1996.

\bibitem{b22} K. Hosnagar, J. Chuang, R. Krishnan and M. Smith, "Service Adoption and Pricing on Content Delivery Network (CDN) Services", Management Science, no. 9, pp 1579-1593, 2008.

\bibitem{b23} P. Maille and B. Tuffin, "Impact of Content Delivery Networks on Service and Content Innovation", ACM SIGMETRICS Performance Evaluation Review, vol. 43, no. 3, pp 49-52, 2105.

\bibitem{b24} E. Gourdin, P. Maille, G. Simon and B. Tuffin, "The Economics of CDNs and Their Impact on Service Fairness", J. IEEE Trans. Networks and Service Management, vol. 14, no. 1, pp 22-33, March 2017.

\bibitem{b25} V. K. Adhikari, Y. Gun, F. Hao, V. Hilt, Z.-L. Zhang, M. Varvello and M. Steiner, "Measurement Study of Netflix, Hulu, and a Tale of Three CDNs", IEEE/ACM Trans. Networking, vol. 23, no. 6, pp. 1984- 1997, 2014.

\bibitem{b26} S. Borst, V. Gupta and A. Walid, "Distributed Caching Algorithms for Content Distribution Networks", Proc. IEEE INFOCOM 2010, pp. 1478- 1486. 

\bibitem{b27} S. Hasan, G. S. C. Dovrolis and R. K. Sitaraman, "Trade-offs in Optimizing Cache Deployments in CDNs", Porc. IEEE INFOCOM 2014, pp. 460-468.

\bibitem{b10} D.N. Serpanos, G. Karakostas and W.H. Wolf, "Effective caching of web objects using Zipf's law", Proc. 2000 IEEE International Conf. Multimedia and Expo, vol. 2, IEEE, pp. 727-730, 2000.

\bibitem{b11} R. Edell and P. Varaiya, "Providing Internet access: What we learn from Index", IEEE Network, no. 5, pp. 18-25, 1999.

\bibitem{b12} S.Lanning, D. Mitra, Q. Wang and M. Wright, "Optimal planning for optical transport networks", Phil. Trans. Royal Soc. of London: Mathematical, Physical and Engineering Sciences, vol. 358, no. 1773, pp.2183-2196, 2000.

\bibitem{b13}P. Reichl, B.Tuffin and R. Schulz, "Logarithmic laws in service quality perception: Where microeconomics meets psychophysics and quality of experience", Telecommunication Systems, vol. 52, pp. 587-600, 2013.


\bibitem{b15} A. Rubenstein, "Perfect Equilibrium in a Bargaining Model", Econometrica, 50:97-109, 1982

\bibitem{b16} R. Gibbons, "Game Theory for Applied Economists", Princeton University Press, 1992

\bibitem{b17} K. Chatterjee, B.Dutta, D. Ray and K.Sengupta, "A Noncooperative Theory of Coalitional Bargaining", Review of Economic Studies, 60(2): 463-477, 1993.

\bibitem{b19} O. Compte and P. Jehiel, "The Coalitional Nash Bargaining Solution", Econometrica, 78(5):1593-1623, 2010.

\bibitem{b21} M.O. Jackson, "Social and Economic Networks", Princeton University Press, 2008.

\bibitem{b28} R.J. Gilbert and D.M.G. Newbery, "Preemptive Patenting and the Persistence of Monopolies", The American Economic Review, 72 (3), 514-526, 1982

\bibitem{b29} K. Arrow, "Economic Welfare and the Allocation of Resources for Inventions," in R. Nelson, Ed., The Rate and Direction of Inventive Activity: Economic and Social Factors, Princeton University Press, pp. 609-626, 1962

\end{thebibliography}
\end{document}